\documentclass{article}
\usepackage{spconf}
\usepackage{amsmath}
\usepackage{graphicx}  
\usepackage{amssymb}
\usepackage{multirow}
\usepackage{threeparttable}
\usepackage{cite}
\usepackage{booktabs}
\usepackage{makecell}
\usepackage{diagbox}
\usepackage{color}

\def\Vec#1{\textit{\boldmath $#1$}}

\def\red#1{#1}

\title{ACCDOA: Activity-Coupled Cartesian Direction of Arrival Representation for Sound Event Localization and Detection}

\makeatletter
\def\@name{ \emph{Kazuki Shimada, Yuichiro Koyama, Naoya Takahashi, Shusuke Takahashi, Yuki Mitsufuji}}
\makeatother

\address{Sony Corporation, Tokyo, Japan}
\begin{document}
\ninept

\maketitle

\begin{abstract}

Neural-network~(NN)-based methods show high performance in sound event localization and detection~(SELD).
Conventional NN-based methods use two branches for a sound event detection~(SED) target and a direction-of-arrival~(DOA) target.
The two-branch representation with a single network has to decide how to balance the two objectives during optimization.
Using two networks dedicated to each task increases system complexity and network size.
To address these problems, we propose an activity-coupled Cartesian DOA~(ACCDOA) representation, which assigns a sound event activity to the length of a corresponding Cartesian DOA vector.
The ACCDOA representation enables us to solve a SELD task with a single target and has two advantages: avoiding the necessity of balancing the objectives and model size increase.
In experimental evaluations with the DCASE 2020 Task 3 dataset, the ACCDOA representation outperformed the two-branch representation in SELD metrics with a smaller network size.
The ACCDOA-based SELD system also performed better than state-of-the-art SELD systems in terms of localization and location-dependent detection.

\end{abstract}

\begin{keywords}
Sound event localization and detection, neural-network
\end{keywords}

\section{Introduction}
\label{sec:intro}

Sound event localization and detection~(SELD) involves
 identifying both the direction-of-arrival~(DOA) and the type of sound events.
SELD has played an essential role in many applications,
 such as surveillance~\cite{crocco2016audio,grobler2017sound,valenzise2007scream},
 bio-diversity monitoring~\cite{chu2009environmental},
 and context-aware devices~\cite{takeda2016sound,yalta2017sound}.
Similar to sound event detection~(SED) and DOA estimation, recent competitions such as DCASE challenge show significant progress in the SELD research area using neural-network (NN)-based methods~\cite{adavanne2019multi,politis2020overview}.

NN-based SELD methods can be categorized into two approaches.
The first aims to
 integrate an NN-based SED method and a physics-based DOA estimation method~\cite{nguyen2020sequence,nguyen2020dcase,yasuda2020sound,xue2019multi}.
The second, which uses only NN-based methods~\cite{adavanne2018sound,politis2020dataset,wang2020ustc,comminiello2019quaternion,phan2020multitask,cao2019polyphonic,kapka2019sound,zhang2019data,cao2020event},
performs excellently when labeled data are available~\cite{adavanne2019multi,politis2020overview}.
Adavanne {\it et al.} proposed SELDnet, which uses two branches for two targets:
 \red{an} SED target and a DOA target~\cite{adavanne2018sound,politis2020dataset}.
Using the two-branch representation, SELDnet aims to solve the multi objectives with a single network.
Cao {\it et al.} proposed a two-stage method that uses two networks: \red{an} SED network and a DOA estimation network~\cite{cao2019polyphonic}.
The two-stage method uses the two single networks to concentrate on each task.

The two-branch representation with a single network has to decide the balance of the objectives.
Such a multi-objective problem is commonly solved by linearly combining the losses with weights defining the trade-off between the loss terms.
However, the weights' values can strongly affect the performance of the model, and their tuning can be cumbersome~\cite{dosovitskiy2020you}.
Although decomposing a network into two task-specific networks avoids the multi-objective problem,
 it also increases the system complexity and the network size.
Keeping the network size small is important for edge devices.

\begin{figure}[t]
    \centering
    \centerline{\includegraphics[width=0.99\linewidth]{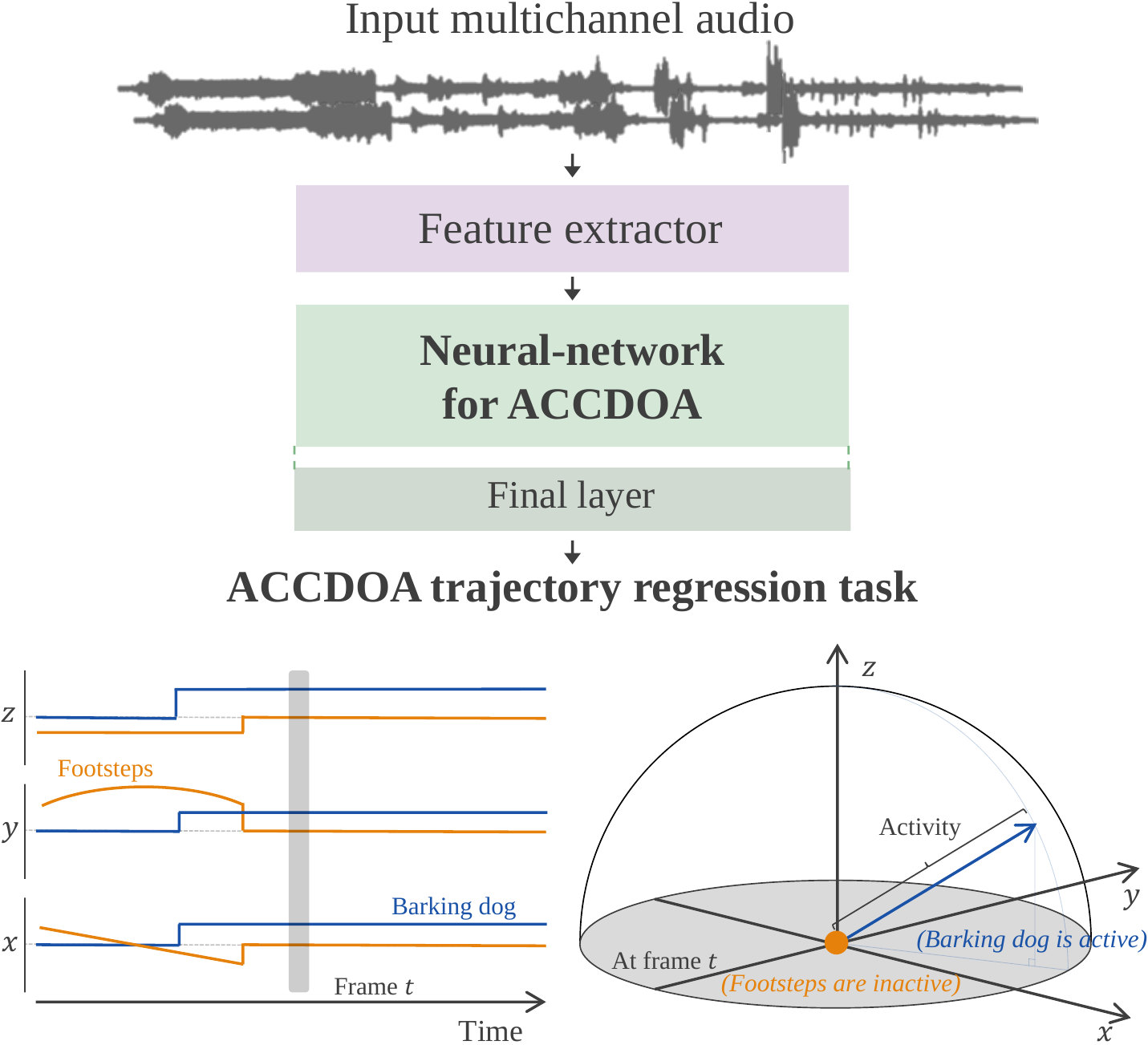}}
    \caption{To solve a SELD task with a single target, after the feature extraction, the network outputs frame-wise ACCDOA vectors for target sound events.
    The ACCDOA vector for each event assigns the activity to the length of the corresponding Cartesian DOA vector.
    When footsteps are inactive, and a barking dog is active at frame $t$, we get two ACCDOA vectors, as shown in the figure's lower right.}
    \label{fig:overview}
\end{figure}

To address the multi-objective problem while keeping the network size small, in this paper, we propose an activity-coupled Cartesian DOA~(ACCDOA) representation for SELD.
The ACCDOA representation assigns a sound event activity to the length of the corresponding Cartesian DOA vector,
 which enables us to handle a SELD task as a single task with a single network.
The ACCDOA representation has two advantages: avoiding the multi-objective problem and network size increase.
A schematic flow of the ACCDOA-based SELD system is shown in Fig.~\ref{fig:overview}.
After the feature extraction, the network outputs frame-wise ACCDOA vectors for target sound events.
The model is trained to minimize the distance between the estimated and the target coordinates in the ACCDOA representation.
When the target indicates no event, then the loss function is calculated only for the activity, not the DOA.
In evaluations using the DCASE 2020 Task 3 dataset~\cite{politis2020dataset},
 the ACCDOA representation outperformed the two-branch representation in SELD metrics with fewer parameters.
The ACCDOA-based SELD system also performed better than state-of-the-art SELD systems in terms of localization error and location-dependent F-score.

\section{Related work}
\label{sec:related}

\begin{figure}[t]
    \centering
    \centerline{\includegraphics[width=0.88\linewidth]{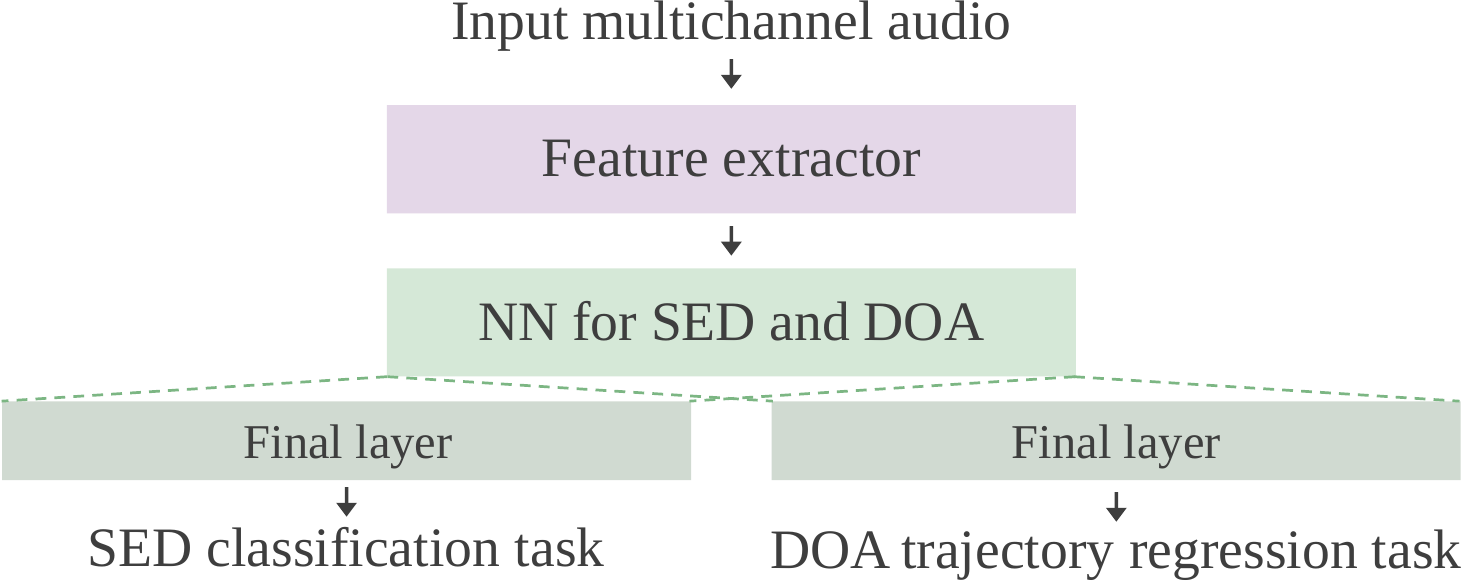}}
    \caption{SELDnet uses an SED target and a DOA target with a single network, which is trained to minimize a combined loss function.}
    \label{fig:seldnet}
\end{figure}

We show an overview of the NN-based SELD methods~\cite{nguyen2020sequence,nguyen2020dcase,yasuda2020sound,xue2019multi,adavanne2018sound,politis2020dataset,wang2020ustc,comminiello2019quaternion,phan2020multitask,cao2019polyphonic,kapka2019sound,zhang2019data,cao2020event},
 which can be categorized into two types of approaches.
After that, we describe two methods closely related to the proposed method: SELDnet~\cite{adavanne2018sound,politis2020dataset} and the two-stage method~\cite{cao2019polyphonic}.

\subsection{Overview of NN-based SELD methods}
\label{ssec:overview}

First, we introduce approaches that
 integrate an NN-based SED method and a physics-based DOA estimation method~\cite{nguyen2020sequence,nguyen2020dcase,yasuda2020sound,xue2019multi}.
Nguyen {\it et al.} used a convolutional recurrent neural-network~(CRNN) for SED
 and a single-source histogram algorithm for DOA estimation
and then integrated their results with a sequence matching network~\cite{nguyen2020sequence,nguyen2020dcase}.
Yasuda {\it et al.} refined an intensity-vector-based DOA estimation method
 by using NN-based denoising and source separation~\cite{yasuda2020sound}.

Second, we introduce approaches using only NN-based methods~\cite{adavanne2018sound,politis2020dataset,wang2020ustc,comminiello2019quaternion,phan2020multitask,cao2019polyphonic,kapka2019sound,zhang2019data,cao2020event}.
Recently, NN-based DOA estimation methods have performed outstandingly
 when labeled data have been available~\cite{tang2019regression}.
Also, this approach offers high performance in the SELD research area
 when we can use labeled data~\cite{adavanne2019multi,politis2020overview}.

Many NN-based methods have tackled a SELD task using a single network with multi objectives~\cite{adavanne2018sound,politis2020dataset,wang2020ustc,comminiello2019quaternion,phan2020multitask}.
Adavanne {\it et al.} proposed SELDnet, which uses two targets: an SED target and a DOA target~\cite{adavanne2018sound,politis2020dataset}.
The system of Wang {\it et al.} performed the best in the DCASE 2020 Task 3 competition,
 and they also used a SELDnet-like network architecture that
 uses an SED target and a DOA target~\cite{wang2020ustc}.

A number of NN-based methods decomposed a SELD task into several subtasks and used subtask-specific networks~\cite{cao2019polyphonic,kapka2019sound,zhang2019data,cao2020event}.
Cao {\it et al.} proposed a two-stage method that uses two networks, i.e., an SED network and a DOA estimation network~\cite{cao2019polyphonic}.
The system of Kapka and Lewandowski performed the best in the DCASE 2019 Task 3,
 and they also decomposed the SELD task into four tasks:
 estimating the number of active sources,
 estimating the DOA of a sound event when there is one active sound source,
 estimating the DOA when there are two, and
 classifying sound events with multi-labels~\cite{kapka2019sound}.

\subsection{SELDnet}
\label{ssec:seldnet}

We show the details of SELDnet~\cite{adavanne2018sound},
 which is the first method that addresses the problem of localizing and recognizing
 more than two overlapping sound events simultaneously
 and tracking their activity per time frame.
After the feature extraction,
 they use a CRNN architecture as an embedding network.
As shown in Fig.~\ref{fig:seldnet},
 this is followed by two branches of fully-connected layers in parallel.
One branch is for SED, and the other is for DOA estimation for each event class.
Some papers show that
 using a Cartesian DOA format is better
 than using a spherical DOA format, i.e., azimuth and elevation,
 owing to the continuity~\cite{adavanne2018sound,tang2019regression}.
A binary-cross-entropy~(BCE) loss is used
 between the SED predictions of SELDnet and reference sound class activities,
 while a mean squared error (MSE) loss is used
 for the DOA estimates of the SELDnet and the reference DOA.
They combine the losses linearly with weights
 and search several weights to balance the losses during optimization.

\subsection{Two-stage method}
\label{ssec:two-stage}

\begin{figure}[t]
    \centering
    \centerline{\includegraphics[width=0.88\linewidth]{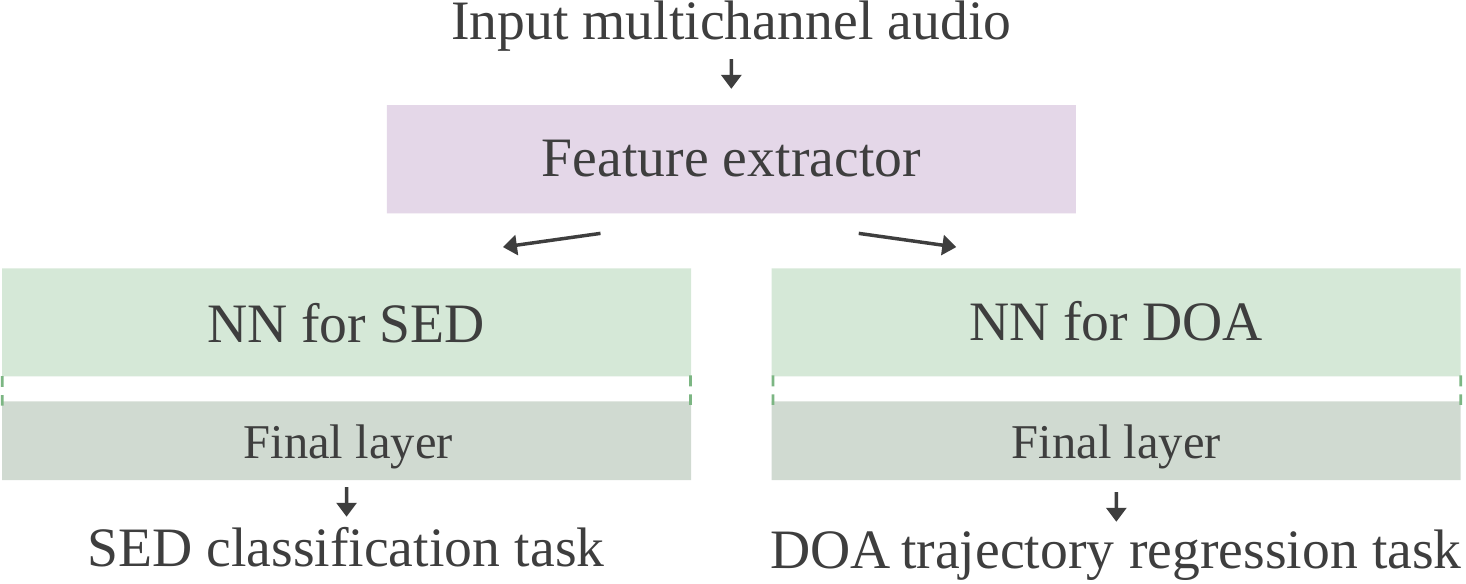}}
    \caption{The two-stage method uses two networks: an SED network and a DOA estimation network.}
    \label{fig:two-stage}
\end{figure}

Fig.~\ref{fig:two-stage} shows the schematic diagram
 of the two-stage method~\cite{cao2019polyphonic},
 which uses two networks: an SED network and a DOA estimation network.
After training only the SED network,
 a part of the network parameters is transferred from the SED network to the DOA estimation network.
Then the DOA estimation network is trained.
A BCE loss is also used
 between the SED network predictions and the reference.
They adopted a masked MSE loss between the DOA estimates and the reference DOA.
When an event is inactive, the MSE loss function is not calculated
 since it is not used in inference time.
SELDnet for the DCASE 2020 Task 3 baseline~\cite{politis2020dataset}
 incorporates the masked MSE loss for its DOA branch.


\section{Proposed Method}
\label{sec:p_method}

We formulate the ACCDOA representation with activity and Cartesian DOA and show how to use the ACCDOA representation for SELD.

\subsection{ACCDOA representation}
\label{ssec:accdoa}

Let $\Vec{a} \in {\mathbb{R}}^{C \times T}$ be $C$-class $T$-frame activities,
 whose reference value of each activity is 1 when an event is active and 0 when inactive\red{, i.e., a reference activity ${a}_{ct}^{*} \in \{0, 1\}$}.
Also let $\Vec{R} \in {\mathbb{R}}^{3 \times C \times T}$ be Cartesian DOAs, where the length of each Cartesian DOA is 1, i.e., $||\Vec{R}_{ct}|| = 1$ when a class $c$ is active.
$||\cdot||$ is Euclidean distance.
Each $C$ sound event class is represented by three nodes
 corresponding to the sound event location in $x$, $y$, and $z$ axes.

\begin{figure}[t]
    \centering
    \centerline{\includegraphics[width=0.88\linewidth]{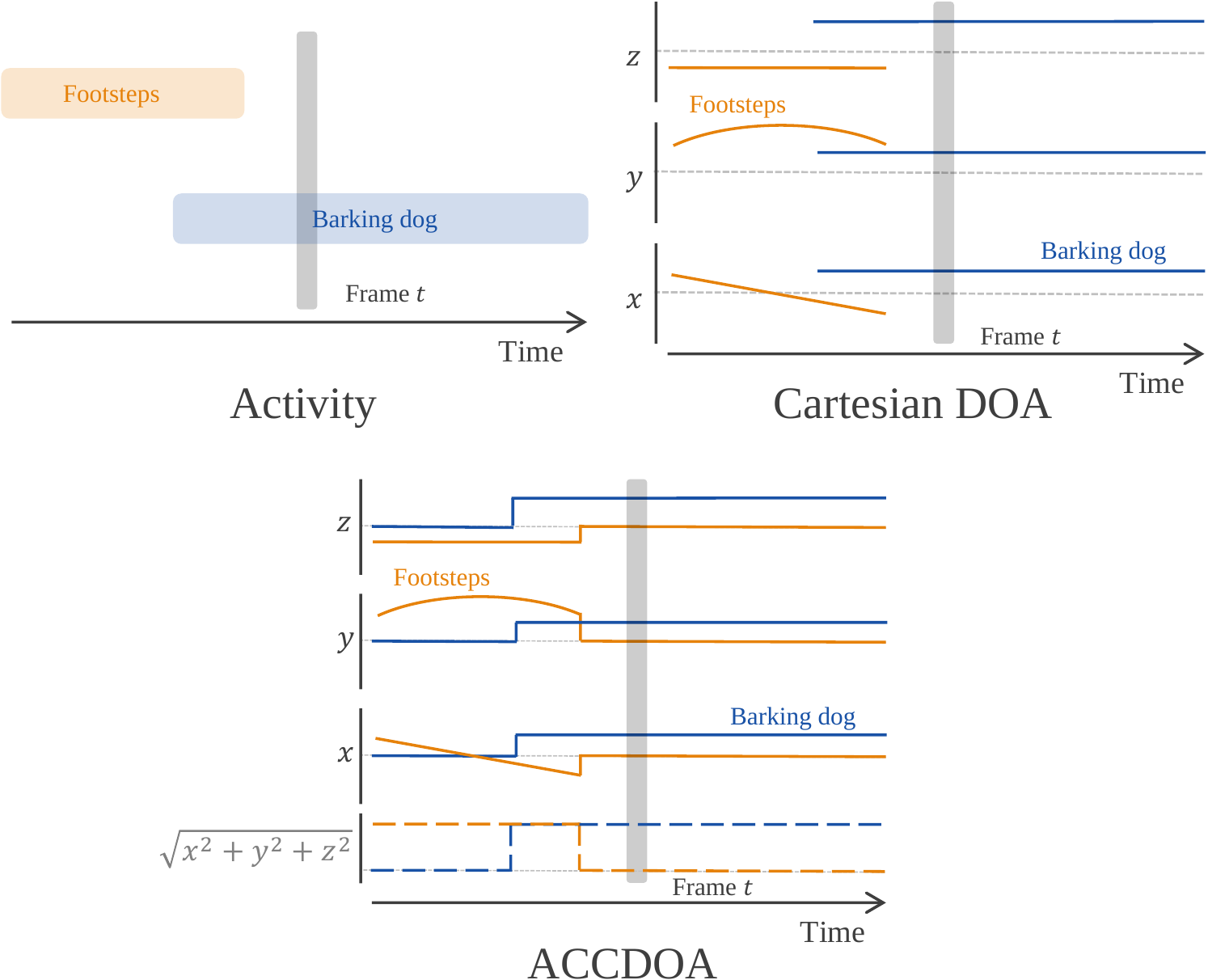}}
    \caption{
    The activity of each event class is represented by the length of the corresponding ACCDOA vector.
    When an event class is active, the DOA is also represented by the direction of the corresponding ACCDOA vector.}
    \label{fig:rep}
    \vspace{-1mm}  
\end{figure}

Then we formulate the ACCDOA representation $\Vec{P} \in {\mathbb{R}}^{3 \times C \times T}$
 with activity and Cartesian DOA:
\begin{align}
    \Vec{P}_{ct} =
        \red{{a}_{ct} \Vec{R}_{ct}.}
    \label{eq:formulate}
\end{align}
Fig.~\ref{fig:rep} shows an example of the relationship between these representations.
Also, activity and Cartesian DOA are obtained from the ACCDOA representation \red{as follows:}
\begin{align}
    {a}_{ct} &= ||\Vec{P}_{ct}||, \\
    \Vec{R}_{ct} &= \frac{\Vec{P}_{ct}}{||\Vec{P}_{ct}||} \:\:\:\: (c {\rm \: is \: active}),
    \label{eq:transform}
\end{align}
where a Cartesian DOA is considered only when the event class is active.
When estimating, we use threshold processing to determine whether an event class is active or not.

\subsection{SELD with ACCDOA representation}
\label{ssec:flow_accdoa}

We use a CRNN architecture as an embedding network, $\mathcal{F}_{\mathrm{embed},{\theta}'}(\cdot)$,
 where ${\theta}'$ is the network parameters.
Let $\Vec{X} \in {\mathbb{R}}^{M \times F \times T'}$ be an input
 consisting of $M$-channel $F$-dimensional $T'$-frame acoustic features.
\begin{align}
    \Vec{E} = \mathcal{F}_{\mathrm{embed},{\theta}'}(\Vec{X}),
    \label{eq:embedding}
\end{align}
where $\Vec{E} \in {\mathbb{R}}^{K \times T}$
 is an embedding vector for the input, $K$ is the number of dimensions of the embedding space.

Using ACCDOA, we can define a SELD task as an ACCDOA estimation problem.
The embedding vector is followed by only one fully-connected layer to estimate ACCDOA representation vectors, $\mathcal{F}_{\mathrm{ACCDOA},{\theta}}(\cdot)$,
 where ${\theta}$ is the parameters.
\begin{align}
    \hat{\Vec{P}}_{t} = \mathcal{F}_{\mathrm{ACCDOA},{\theta}}(\Vec{E}_{t}),
    \label{eq:final_accdoa}
\end{align}
\red{where $\hat{\Vec{P}}$ is an estimated ACCDOA representation.}
We estimate SELD outputs without two task-specific branches or two task-specific networks.
That leads to simple network architecture and small network size.

\begin{figure}[t]
    \centering
    \centerline{\includegraphics[width=0.90\linewidth]{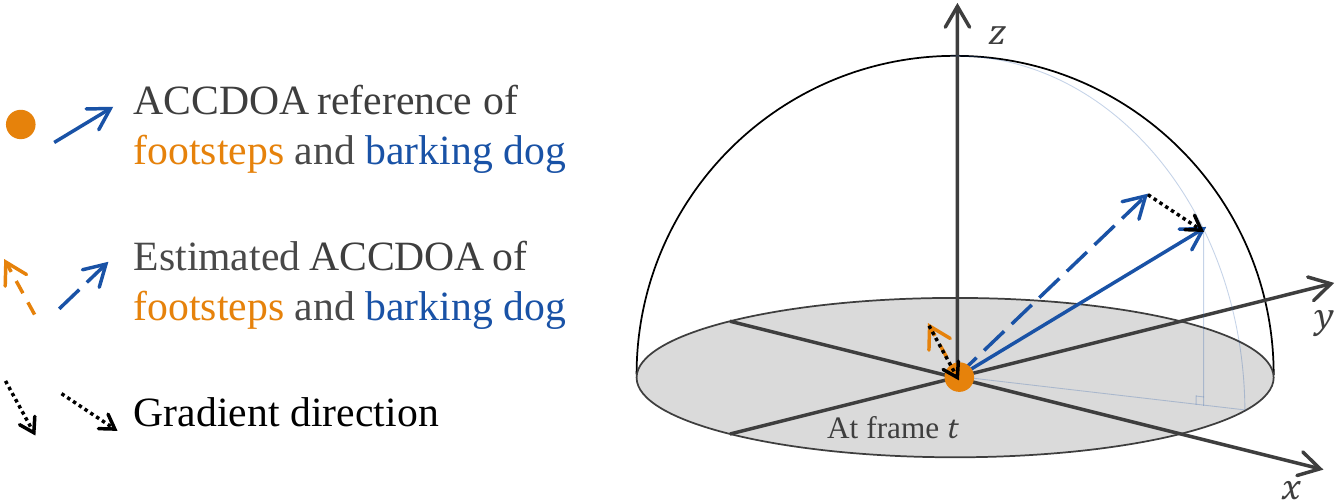}}
    \caption{\red{ACCDOA vectors at frame $t$ in Fig.~\ref{fig:rep}.
    A Solid line or a dot shows an ACCDOA reference, whose length is 1 when an event is active and 0 when inactive.
    Dashed lines show ACCDOA estimated by a network, which is trained to bring the estimation closer to the reference.
    Dotted lines depict the gradient directions.}}
    \label{fig:target}
\end{figure}

Fig.~\ref{fig:target} shows an example of the differences
 between reference and estimated ACCDOA at frame $t$ in Fig.~\ref{fig:rep}.
We use MSE as a single loss function for a SELD task.
\begin{align}
    {l}_\mathrm{ACCDOA} = \frac{1}{CT} \sum_{c}^{C} \sum_{t}^{T} \mathrm{MSE}(\red{\Vec{P}_{ct}^{*}}, \hat{\Vec{P}}_{ct}),
    \label{eq:loss_accdoa}
\end{align}
\red{where $\Vec{P}^{*}$ is a reference ACCDOA representation.}
When the ground truth indicates no event, then the MSE loss function is calculated only for the activity, not the DOA.
In the original SELDnet~\cite{adavanne2018sound}, even in the case of no event, DOA loss needed to be calculated.
Using ACCDOA can avoid such calculations, as can using masked MSE.
Activity is no longer interpreted as a posterior probability because BCE is not used.
However, threshold processing can be performed in the same manner.
This single loss enables us to avoid the multi-objective problem and focus on other hyper-parameter tunings.

\section{Experimental evaluation}
\label{sec:exp}

\begin{table*}[t]
    \centering
    \caption{SELD performance with two types of CRNNs in different representations evaluated for the DCASE 2020 Task 3 development set.}
    \vspace{0.5mm}
    \scalebox{0.87}{
        \begin{tabular}{ll|c|cccc|cccc}
        \toprule
        & & Number of & \multicolumn{4}{c|}{Validation split} & \multicolumn{4}{c}{Testing split} \\
        Network & Representation & parameters & ${LE}_{CD}$ & ${LR}_{CD}$ & ${ER}_{20^{\circ}}$ & ${F}_{20^{\circ}}$ & ${LE}_{CD}$ & ${LR}_{CD}$ & ${ER}_{20^{\circ}}$ & ${F}_{20^{\circ}}$ \\
        \midrule
        CRNN & Two-branch for SELDnet   & 5.24 M & $9.6^{\circ}$ & 78.7 & 0.38 & 72.5 & $11.0^{\circ}$ & 70.9 & 0.45 & 64.1 \\
             & Two-branch for Two-stage & 9.41 M & $8.9^{\circ}$ & {\bf 79.5} & {\bf 0.36} & {\bf 74.3} & $10.9^{\circ}$ & {\bf 73.0} & 0.44 & 66.3 \\
             & ACCDOA (proposed)        & {\bf 4.71 M} & ${\bf 8.5^{\circ}}$ & 76.0 & 0.39 & 72.4 & ${\bf 9.6^{\circ}}$ & 71.9 & {\bf 0.43} & {\bf 67.5} \\
        \midrule
        RD3Net & Two-branch for SELDnet   & 1.62 M & $9.2^{\circ}$ & 83.1 & 0.33 & 77.1 & $11.6^{\circ}$ & {\bf 79.7} & 0.38 & 70.7 \\
              & Two-branch for Two-stage & 3.12 M & $8.0^{\circ}$ & 83.0 & 0.32 & 78.5 & $9.7^{\circ}$ & 76.2 & 0.39 & 70.5 \\
              & ACCDOA (proposed)        & {\bf 1.56 M} & ${\bf 7.6^{\circ}}$ & {\bf 83.6} & {\bf 0.28} & {\bf 80.9} & ${\bf 8.3^{\circ}}$ & 77.5 & {\bf 0.36} & {\bf 74.2}  \\
        \bottomrule
        \end{tabular}
    }
    \label{tb:result}
    \vspace{-5.5mm}
\end{table*}

We evaluated the ACCDOA representation and the two-branch representation
 using TAU Spatial Sound Events 2020
 with the setup for DCASE 2020 Task 3~\cite{politis2020dataset}.
We also compared our proposed method
 with state-of-the-art SELD systems without model ensembles.

\subsection{Task setups}
\label{ssec:tasks}

We used the development set of TAU Spatial Sound Events 2020 - Ambisonic
 with the suggested setup for DCASE 2020 Task 3~\cite{politis2020dataset}.
The dataset contained 600 one-minute sound scene recordings: 400 for training, 100 for validation, and 100 for testing.
The sound scene recordings were synthesized by adding sound event samples convolved with room impulse response~(RIR) to spatial ambient noise.
The sound event samples were from the NIGENS general sound events database~\cite{trowitzsch2019nigens}, which consisted of 14 event classes such as footsteps and a barking dog.
The RIRs and ambient noise recordings were collected at 15 different indoor locations.
Each event had an equal probability of being either static or moving.
The moving sound events were synthesized with 10, 20, or 40 degrees per second.
Signal-to-noise ratios ranged from 6 dB to 30 dB.

Following the setup, four metrics were used for the evaluation~\cite{mesaros2019joint}.
The first was the localization error ${LE}_{CD}$, which expresses the average angular distance between the same class's predictions and references.
The second was a simple localization recall metric ${LR}_{CD}$, which tells the true positive rate of how many of these localization estimates were detected in a class out of the total number of class instances.
The next two metrics were the location-dependent error rate ${ER}_{20^{\circ}}$ and F-score ${F}_{20^{\circ}}$,
 where predictions were considered true positives only when the distance from the reference was less than  $20^{\circ}$.

\subsection{Experimental settings}
\label{ssec:system}


We compared the ACCDOA representation
 with the two-branch representations for SELDnet (Two-branch for SELDnet) and the two-stage method (Two-branch for Two-stage).
The differences between ACCDOA and Two-branch for SELDnet were only after the embedding vector.
Following the baseline for DCASE 2020 Task 3, each task-specific branch for SELDnet had two fully-connected layers, and the loss functions were BCE and masked MSE.
The loss weight value was 10, chosen from 1, 5, 10, 20, 100.
Also, the differences between ACCDOA and Two-branch for Two-stage were the number of networks.
\red{In both representations,} other system configurations \red{such as feature extraction, data augmentation, and post-processing} mostly followed our DCASE 2020 challenge settings~\cite{shimada2020sound}.

Multichannel amplitude spectrograms and inter-channel phase differences~(IPDs)
 were used as frame-wise features.
Since the input consists of four-channel signals,
 we extracted four amplitude spectrograms and three IPDs.

Following Takahashi {\it et al.}~\cite{takahashi2016deep,takahashi2017aenet},
 we applied the equalized mixture data augmentation~(EMDA) method where up to two sound events are mixed
 with random amplitudes, delays, and modulations of frequency characteristics, i.e., equalization.
We also adopted the spatial augmentation method of Mazzon {\it et al.}~\cite{mazzon2019first}, which rotates the training data represented in the first-order Ambisonic~(FOA) format.
We used the multichannel version of SpecAugment~\cite{shimada2020sound}.
In addition to the time-frequency hard-masking schemes~\cite{park2019specaugment} applied to amplitude spectrograms, the masking schemes were extended to the channel dimension~\cite{shimada2020sound}.
The target channel for the channel masking,
 $m_{0}$, was chosen from $[0, M)$ where $M$ denotes the number of microphone channels.
For the IPD features, instead of multiplying a mask value by the original value,
 the actual values were replaced with random values,
 where the values were sampled from a uniform distribution ranging from 0 to $2\pi$.

We prepared two types of CRNNs:
a conventional CRNN used in~\cite{cao2019polyphonic} and RD3Net~\cite{takahashi2020d3net}.
RD3Net has shown state-of-the-art performance in music source separation\red{, and the advantage is the efficiency of modeling a large receptive field, which is beneficial for many tasks that require a long context for an accurate prediction~\cite{takahashi2020d3net}.}
The adaptation to the SELD task includes three modifications.
First, we omitted dense blocks in the up-sampling path
 because high frame-rate prediction is unnecessary for the SELD task.
Second, we placed gated recurrent unit~(GRU) cells existing only in the bottleneck part.
Third, the batch normalization was replaced with the network deconvolution~\cite{ye2020network}.
In each dense block, the initial convolution's dilation factor was set to one,
 and the dilation factor doubled every time the next convolution was applied.

We split the 60-second input audio into shorter segments with overlap during the inference,
 processed each segment,
 and averaged the results of overlapped frames.
To further improve the performance,
 we conducted post-processing with the following procedure:
 rotating the FOA data, estimating the ACCDOA vectors, rotating the vectors back,
 and averaging the vectors of different rotation patterns.

The sampling frequency was set to 24 kHz.
The short-term Fourier transform~(STFT) was applied with a configuration of 20 ms frame length and 10 ms frame hop.
The frame length of the network input was 128 frames.
During the inference time, the frame shift length was set to 20 frames.
We used a batch size of 32, and each training sample was generated on-the-fly~\cite{erdogan2018investigations}.
The learning rate was set to 0.001 and decayed 0.9 times every 20,000 iterations.
We used the Adam optimizer with a weight decay of~$10^{-6}$.

We also compared our proposed method with state-of-the-art SELD systems without model ensembles.
In this case, to further improve the performance, we used RD3Net and extended the network input's frame length to 1,024 frames.

\subsection{Experimental results}
\label{ssec:res}

\begin{table}[t]
    \centering
    \caption{SELD performance of state-of-the-art systems and our ACCDOA-based system for the development set. \red{PP denotes post-processing described in Sec. 4.2.}}
    \vspace{0.5mm}
    \scalebox{0.87}{
        \begin{tabular}{l|c|cccc}
        \toprule
        & \red{\# of} & \multicolumn{4}{c}{Testing split} \\
        System & \red{params} & ${LE}_{CD}$ & ${LR}_{CD}$ & ${ER}_{20^{\circ}}$ & ${F}_{20^{\circ}}$ \\
        \midrule
        Wang's~\cite{wang2020ustc}              & N/A       & $9.4^{\circ}$ & {\bf 82.8} & {\bf 0.29} & 76.4 \\
        Nguyen's~\cite{nguyen2020dcase}         & 2.28 M    & $13.5^{\circ}$ & 81.5 & 0.38 & 69.4 \\
        FOA Baseline~\cite{politis2020dataset}  & 0.51 M    & $22.8^{\circ}$ & 60.7 & 0.72 & 37.4 \\
        \midrule
        \red{Ours w/o PP}                       & 1.56 M    & $10.2^{\circ}$ & 79.1 & 0.36 & 73.0 \\
        Ours                                    & 1.56 M    & ${\bf 7.9^{\circ}}$ & 80.5 & 0.32 & {\bf 76.8} \\
        \bottomrule
        \end{tabular}
    }
    \label{tb:result_sys}
    \vspace{-5.5mm}
\end{table}

Table~\ref{tb:result} shows the performance with two types of CRNNs in different representations.
As shown in the table, the ACCDOA representation has fewer parameters than two-branch representations.
The ACCDOA representation outperformed the two-branch representation for SELDnet
 for most metrics with both networks.
While the ACCDOA representation showed 1.0 point higher ${LR}_{CD}$
 than the Two-branch for SELDnet in the testing split with CRNN,
 the ACCDOA representation improved ${F}_{20^{\circ}}$ by 3.4 points.
This result suggests that the ACCDOA representation is more effective in improving location-dependent detection.
Our experimental results show that the ACCDOA representation can simultaneously focus on estimating activity and DOA with a single target using a single network.

Table~\ref{tb:result_sys} shows the performances of the SELD systems without model ensembles.
Our ACCDOA-based system performed the best in terms of ${LE}_{CD}$ and ${F}_{20^{\circ}}$ for the DCASE 2020 Task 3 development set \red{with small number of parameters.}
\red{The results also show the post-processing especially improves localization performance.}

The dataset used in the experiments can contain up to two overlapping sound event classes, and almost all overlaps are those of different classes.
In the future, when we need to tackle overlaps of the same class, a promising way is to incorporate a multi-track extension~\cite{nguyen2020sequence,nguyen2020dcase,cao2020event} to the ACCDOA representation.

\section{Conclusion}
\label{sec:conclusion}

We proposed the activity-coupled Cartesian direction-of-arrival~(ACCDOA) representation, which assigns a sound event activity to the length of a corresponding Cartesian DOA vector for sound event localization and detection~(SELD).
The ACCDOA representation enables us to solve a SELD task with a single target using a single network.
The ACCDOA representation avoids the necessity of balancing objectives for sound event detection~(SED) and DOA estimation and the increase of network size.
In the evaluations on the SELD task for DCASE 2020 Task 3, the ACCDOA representation outperformed the conventional two-branch representations in the joint SELD metrics with fewer parameters.
The ACCDOA-based system also performed better than state-of-the-art systems in terms of localization error and location-dependent F-score.

\bibliographystyle{IEEEtran}
\bibliography{refs}

\begin{thebibliography}{10}
\providecommand{\url}[1]{#1}
\csname url@samestyle\endcsname
\providecommand{\newblock}{\relax}
\providecommand{\bibinfo}[2]{#2}
\providecommand{\BIBentrySTDinterwordspacing}{\spaceskip=0pt\relax}
\providecommand{\BIBentryALTinterwordstretchfactor}{4}
\providecommand{\BIBentryALTinterwordspacing}{\spaceskip=\fontdimen2\font plus
\BIBentryALTinterwordstretchfactor\fontdimen3\font minus
  \fontdimen4\font\relax}
\providecommand{\BIBforeignlanguage}[2]{{%
\expandafter\ifx\csname l@#1\endcsname\relax
\typeout{** WARNING: IEEEtran.bst: No hyphenation pattern has been}%
\typeout{** loaded for the language `#1'. Using the pattern for}%
\typeout{** the default language instead.}%
\else
\language=\csname l@#1\endcsname
\fi
#2}}
\providecommand{\BIBdecl}{\relax}
\BIBdecl

\bibitem{crocco2016audio}
M.~Crocco, M.~Cristani, A.~Trucco, and V.~Murino, ``Audio surveillance: A
  systematic review,'' \emph{ACM Computing Surveys}, vol.~48, no.~4, pp. 1--46,
  2016.

\bibitem{grobler2017sound}
C.~Grobler, C.~P. Kruger, B.~J. Silva, and G.~P. Hancke, ``Sound based
  localization and identification in industrial environments,'' in \emph{Proc.
  of IEEE IECON}, 2017, pp. 6119--6124.

\bibitem{valenzise2007scream}
G.~Valenzise, L.~Gerosa, M.~Tagliasacchi, F.~Antonacci, and A.~Sarti, ``Scream
  and gunshot detection and localization for audio-surveillance systems,'' in
  \emph{Proc. of IEEE AVSS}, 2007, pp. 21--26.

\bibitem{chu2009environmental}
S.~Chu, S.~Narayanan, and C.-C.~J. Kuo, ``Environmental sound recognition with
  time--frequency audio features,'' \emph{IEEE Trans. on ASLP}, vol.~17, no.~6,
  pp. 1142--1158, 2009.

\bibitem{takeda2016sound}
R.~Takeda and K.~Komatani, ``Sound source localization based on deep neural
  networks with directional activate function exploiting phase information,''
  in \emph{Proc. of IEEE ICASSP}, 2016, pp. 405--409.

\bibitem{yalta2017sound}
N.~Yalta, K.~Nakadai, and T.~Ogata, ``Sound source localization using deep
  learning models,'' \emph{Journal of Robotics and Mechatronics}, vol.~29,
  no.~1, pp. 37--48, 2017.

\bibitem{adavanne2019multi}
S.~Adavanne, A.~Politis, and T.~Virtanen, ``A multi-room reverberant dataset
  for sound event localization and detection,'' in \emph{Proc. of DCASE
  workshop}, 2019.

\bibitem{politis2020overview}
A.~Politis, A.~Mesaros, S.~Adavanne, T.~Heittola, and T.~Virtanen, ``Overview
  and evaluation of sound event localization and detection in {DCASE} 2019,''
  \emph{arXiv:2009.02792}, 2020.

\bibitem{nguyen2020sequence}
T.~N.~T. Nguyen, D.~L. Jones, and W.-S. Gan, ``A sequence matching network for
  polyphonic sound event localization and detection,'' in \emph{Proc. of IEEE
  ICASSP}, 2020, pp. 71--75.

\bibitem{nguyen2020dcase}
------, ``{DCASE 2020 Task 3}: Ensemble of sequence matching networks for
  dynamic sound event localization, detection, and tracking,'' in \emph{Tech.
  report of DCASE Challange}, 2020.

\bibitem{yasuda2020sound}
M.~Yasuda, Y.~Koizumi, S.~Saito, H.~Uematsu, and K.~Imoto, ``Sound event
  localization based on sound intensity vector refined by {DNN}-based denoising
  and source separation,'' in \emph{Proc. of IEEE ICASSP}, 2020, pp. 651--655.

\bibitem{xue2019multi}
W.~Xue, T.~Ying, Z.~Chao, and D.~Guohong, ``Multi-beam and multi-task learning
  for joint sound event detection and localization,'' in \emph{Tech. report of
  DCASE Challange}, 2019.

\bibitem{adavanne2018sound}
S.~Adavanne, A.~Politis, J.~Nikunen, and T.~Virtanen, ``Sound event
  localization and detection of overlapping sources using convolutional
  recurrent neural networks,'' \emph{IEEE JSTSP}, vol.~13, no.~1, pp. 34--48,
  2018.

\bibitem{politis2020dataset}
A.~Politis, S.~Adavanne, and T.~Virtanen, ``A dataset of reverberant spatial
  sound scenes with moving sources for sound event localization and
  detection,'' \emph{arXiv:2006.01919}, 2020.

\bibitem{wang2020ustc}
Q.~Wang, H.~Wu, Z.~Jing, F.~Ma, Y.~Fang, Y.~Wang, T.~Chen, J.~Pan, J.~Du, and
  C.-H. Lee, ``The {USTC-iFLYTEK} system for sound event localization and
  detection of {DCASE2020} challenge,'' in \emph{Tech. report of DCASE
  Challange}, 2020.

\bibitem{comminiello2019quaternion}
D.~Comminiello, M.~Lella, S.~Scardapane, and A.~Uncini, ``Quaternion
  convolutional neural networks for detection and localization of {3D} sound
  events,'' in \emph{Proc. of IEEE ICASSP}, 2019, pp. 8533--8537.

\bibitem{phan2020multitask}
H.~Phan, L.~Pham, P.~Koch, N.~Q. Duong, I.~McLoughlin, and A.~Mertins, ``On
  multitask loss function for audio event detection and localization,''
  \emph{arXiv:2009.05527}, 2020.

\bibitem{cao2019polyphonic}
Y.~Cao, Q.~Kong, T.~Iqbal, F.~An, W.~Wang, and M.~D. Plumbley, ``Polyphonic
  sound event detection and localization using a two-stage strategy,'' in
  \emph{Proc. of DCASE Workshop}, 2019.

\bibitem{kapka2019sound}
S.~Kapka and M.~Lewandowski, ``Sound source detection, localization and
  classification using consecutive ensemble of crnn models,'' in \emph{Proc. of
  DCASE Workshop}, 2019.

\bibitem{zhang2019data}
J.~Zhang, W.~Ding, and L.~He, ``Data augmentation and prior knowledge-based
  regularization for sound event localization and detection,'' in \emph{Tech.
  Report of DCASE Challenge}, 2019.

\bibitem{cao2020event}
Y.~Cao, T.~Iqbal, Q.~Kong, Y.~Zhong, W.~Wang, and M.~D. Plumbley,
  ``Event-independent network for polyphonic sound event localization and
  detection,'' \emph{arXiv:2010.00140}, 2020.

\bibitem{dosovitskiy2020you}
A.~Dosovitskiy and J.~Djolonga, ``{You Only Train Once}: Loss-conditional
  training of deep networks,'' in \emph{Proc. of ICLR}, 2020.

\bibitem{tang2019regression}
Z.~Tang, J.~D. Kanu, K.~Hogan, and D.~Manocha, ``Regression and classification
  for direction-of-arrival estimation with convolutional recurrent neural
  networks,'' in \emph{Proc. of Interspeech}, 2019, pp. 654--658.

\bibitem{trowitzsch2019nigens}
I.~Trowitzsch, J.~Taghia, Y.~Kashef, and K.~Obermayer, ``The {NIGENS} general
  sound events database,'' \emph{arXiv:1902.08314}, 2019.

\bibitem{mesaros2019joint}
A.~Mesaros, S.~Adavanne, A.~Politis, T.~Heittola, and T.~Virtanen, ``Joint
  measurement of localization and detection of sound events,'' in \emph{Proc.
  of IEEE WASPAA}, 2019.

\bibitem{shimada2020sound}
K.~Shimada, N.~Takahashi, S.~Takahashi, and Y.~Mitsufuji, ``Sound event
  localization and detection using activity-coupled {Cartesian DOA} vector and
  {RD3Net},'' in \emph{Tech. report of DCASE Challange}, 2020.

\bibitem{takahashi2016deep}
N.~Takahashi, M.~Gygli, B.~Pfister, and L.~V. Gool, ``Deep convolutional neural
  networks and data augmentation for acoustic event detection,'' in \emph{Proc.
  of Interspeech}, 2016, pp. 2982--2986.

\bibitem{takahashi2017aenet}
N.~Takahashi, M.~Gygli, and L.~{Van Gool}, ``{AENet}: Learning deep audio
  features for video analysis,'' \emph{IEEE Trans. on Multimedia}, vol.~20, pp.
  513--524, 2017.

\bibitem{mazzon2019first}
L.~Mazzon, Y.~Koizumi, M.~Yasuda, and N.~Harada, ``First order ambisonics
  domain spatial augmentation for {DNN}-based direction of arrival
  estimation,'' in \emph{Proc. of DCASE Workshop}, 2019.

\bibitem{park2019specaugment}
D.~S. Park, W.~Chan, Y.~Zhang, C.-C. Chiu, B.~Zoph, E.~D. Cubuk, and Q.~V. Le,
  ``{SpecAugment}: A simple data augmentation method for automatic speech
  recognition,'' in \emph{Proc. of Interspeech}, 2019, pp. 2613--2617.

\bibitem{takahashi2020d3net}
N.~Takahashi and Y.~Mitsufuji, ``{D3Net}: Densely connected multidilated
  densenet for music source separation,'' \emph{arXiv:2010.01733}, 2020.

\bibitem{ye2020network}
C.~Ye, M.~Evanusa, H.~He, A.~Mitrokhin, T.~Goldstein, J.~A. Yorke,
  C.~Fermuller, and Y.~Aloimonos, ``Network deconvolution,'' in \emph{Proc. of
  ICLR}, 2020.

\bibitem{erdogan2018investigations}
H.~Erdogan and T.~Yoshioka, ``Investigations on data augmentation and loss
  functions for deep learning based speech-background separation,'' in
  \emph{Proc. of Interspeech}, 2018, pp. 3499--3503.

\end{thebibliography}

\end{document}